\documentclass[fleqn,usenatbib]{mnras}

\usepackage{newtxtext,newtxmath}
\usepackage[T1]{fontenc}

\DeclareRobustCommand{\VAN}[3]{#2}
\let\VANthebibliography\thebibliography
\def\thebibliography{\DeclareRobustCommand{\VAN}[3]{##3}\VANthebibliography}

\usepackage{graphicx}	% Including figure files
\usepackage{amsmath}	% Advanced maths commands

\usepackage{amssymb}	% Extra maths symbols
\usepackage{cleveref}               % Easy cross-reference naming
\usepackage[dvipsnames]{xcolor}     % Colored text

\usepackage{bm}                     % More math bold fonts
\usepackage{siunitx}
\usepackage{xfrac}
\usepackage{pgfplots}
\DeclareUnicodeCharacter{2212}{−}
\usepgfplotslibrary{groupplots,dateplot}
\usetikzlibrary{patterns,shapes.arrows}
\pgfplotsset{compat=newest}
\usepackage[inline,shortlabels]{enumitem}
\usepackage{lineno}
\graphicspath{{include/}}
\let\oldequation\equation
\let\oldendequation\endequation
\renewenvironment{equation}
  {\linenomathNonumbers\oldequation}
  {\oldendequation\endlinenomath}
\expandafter\let\expandafter\oldequations\csname equation*\endcsname
\expandafter\let\expandafter\oldendequations\csname endequation*\endcsname
\renewenvironment{equation*}
  {\linenomathWithnumbers\oldequations}
  {\oldendequations\endlinenomath}
\let\oldalign\align
\let\oldendalign\endalign
\renewenvironment{align}
  {\linenomathNonumbers\oldalign}
  {\oldendalign\endlinenomath}
\expandafter\let\expandafter\oldaligns\csname align*\endcsname
\expandafter\let\expandafter\oldendaligns\csname endalign*\endcsname
\renewenvironment{align*}
  {\linenomathWithnumbers\oldaligns}
  {\oldendaligns\endlinenomath}
\renewenvironment{eqnarray}{\align}{\endalign}
\expandafter\let\expandafter\oldeqnarrays\csname eqnarray*\endcsname
\expandafter\let\expandafter\oldendeqnarrays\csname eqnarray*\endcsname
\renewenvironment{eqnarray*}{\expandafter\csname align*\endcsname}{\expandafter\csname endalign*\endcsname}

\newcommand{\redmapper}{redMaPPer}

\newcommand{\Redge}{\ensuremath{R_{\rm edge}}}
\newcommand{\orb}{{\rm orb}}
\renewcommand{\inf}{{\rm inf}}
\newcommand{\los}{{\rm los}}
\newcommand{\p}{{\rm p}}
\newcommand{\zp}{z_{\rm p}}

\newcommand{\cl}{{\rm cl}}
\newcommand{\hc}{\ensuremath{H_0}}

\DeclareSIUnit{\sqdeg}{\text{square degrees}}
\DeclareSIUnit{\degr}{\text{deg}}
\DeclareSIUnit{\amin}{\text{arcmin}}
\DeclareSIUnit{\h}{\ensuremath{\mathit{h}}}
\DeclareSIUnit{\parsec}{\text{pc}}
\DeclareSIUnit{\million}{\text{M}}
\DeclareSIUnit{\thousand}{\text{k}}
\makeatletter
\newcommand{\vvec}[1]{\ensuremath{\vec{#1}%
    \@ifnextchar^{\,}{}}%
}
\makeatother

\newcommand{\Romannumeral}[1]{\MakeUppercase{\romannumeral #1}}
\NewDocumentCommand{\mydefcitealias}{ m m }{%
    \defcitealias{#1}{#2}%
    \citeauthor{#1} (\citeyear{#1}; hereafter \citetalias{#1})%
}

\newcommand{\sigmahfid}{0.021}

\newcommand{\sigmahbest}{0.0088}

\Crefname{figure}{Fig.}{Figs.}
\Crefname{equation}{Eqn.}{Eqns.}

\title[Cluster Edge Cosmology]{Measuring Cosmological Distances Using Cluster Edges as a Standard Ruler}

\author[E. L. Wagoner et al.]{
Erika L. Wagoner,$^{1}$\thanks{E-mail: wagoner47@email.arizona.edu (ELW)}
Eduardo Rozo,$^{1}$
Han Aung,$^2$
Daisuke Nagai$^{2,3}$
\\
$^{1}$Department of Physics, University of Arizona, 1118 E. Fourth Street, Tucson, AZ, 85721, USA\\
$^{2}$Department of Physics, Yale University, New Haven, CT 06520, USA\\
$^{3}$Department of Astronomy, Yale University, New Haven, CT 06520, USA
}

\date{Accepted XXX. Received YYY; in original form ZZZ}

\pubyear{2020}

\begin{document}

\label{firstpage}
\pagerange{\pageref{firstpage}--\pageref{lastpage}}
\maketitle

\begin{abstract}
The line-of-sight velocity dispersion profile of galaxy clusters exhibits a ``kink'' corresponding to the spatial extent of orbiting galaxies. Because the spatial extent of a cluster is correlated with the amplitude of the velocity dispersion profile, we can utilise this feature as a gravity-calibrated standard ruler. Specifically, the amplitude of the velocity dispersion data allows us to infer the physical cluster size. Consequently, observations of the angular scale of the ``kink'' in the profile can be translated into a distance measurement to the cluster. Assuming the relation between cluster radius and cluster velocity dispersion can be calibrated from simulations, we forecast that with existing data from the Sloan Digital Sky Survey (SDSS) we will be able to measure the Hubble constant with \SI{3.0}{\percent} precision. Implementing our method with data from the Dark Energy Spectroscopic Instrument (DESI) will result in a \SI{1.3}{\percent} measurement of the Hubble constant. Adding cosmological supernova data improves the uncertainty of the DESI measurement to \SI{0.7}{\percent}.
\end{abstract}


%%%%%%%%%%%%%%%%%%%%%%%%%%%%%%%%%%%%%%%%%%%%%%%%%%

%%%%%%%%%%%%%%%%% BODY OF PAPER %%%%%%%%%%%%%%%%%%

\section{Introduction}

One of the greatest challenges to the standard model of modern cosmology is the $4.4\sigma$ tension between the Hubble constant (\hc{}) inferred from the Cosmic Microwave Background \citep[CMB;][]{2018arXiv180706209P} and that from supernova measurements by the SH0ES team \citep{2019ApJ...876...85R}. This discrepancy could be the first hint of a breakdown of the standard model of cosmology, and has therefore led to a broad range of new theoretical ideas for how this tension can be resolved \citep[for a very nice review see][]{knoxmilleta19}. To unambiguously establish the need for new physics, an independent confirmation of the Hubble tension must be achieved. 

Despite tremendous progress in this regard \citep[for a discussion of recent results see e.g.][]{verdeetal19}, the picture is not yet clear: while the predicted value for the Hubble constant from early-universe physics is quite robustly predicted, the local value of the Hubble constant is less robustly determined. Specifically, an independent distance ladder calibration of the Hubble constant using the tip of the red giant branch method results in a lower Hubble constant value \citep[$h=0.696\pm 0.019$,][]{freedmanetal20}. Distance calibrations via surface brightness fluctuations results in an intermediate Hubble value, consistent with the SH0ES and early-universe estimates \citep[$h=0.705\pm 0.034$,][]{khetanetal20}, while cosmic chronometers also favour low Hubble values \citep[e.g.][]{haridasuetal8,gomezamendola19}. Early strong lensing constraints from the H0LICOW team favoured a high Hubble value \citep[$h=0.733\pm 0.017$,][]{wongetal19}, but these values are sensitive to the detailed radial profile assumed for the lenses \citep{kochanek20}. Allowing for greater freedom in the mass models for the lenses while incorporating galaxy velocity dispersion data results in a low Hubble constant estimate \citep[$h=0.674\pm 0.04$,][]{birreretal20}. On the other hand, megamaser-based geometric distances lead to a high $H_0$ value \citep[$h=0.74\pm 0.03$,][]{pesceetal20}. While gravitational waves may soon be able to resolve this dichotomy, current constraints are still limited by low number statistics \citep{ligo19}. Additional discussion can be found in \citet{2020arXiv200811284D}.

In light of the current state of affairs, any new independent distance estimator is of great interest, provided the measurement can achieve percent level precision. In this paper, we propose one such method. Our method makes use of the cluster edge radius identified by \mydefcitealias{2020arXiv200311555T}{Paper \Romannumeral{1}} and the corresponding dark matter halo edge radius identified by \mydefcitealias{2020arXiv200311557A}{Paper \Romannumeral{2}}. These radii appear to be intimately related to (but are somewhat larger than) the more well known splashback radius \citep[see e.g.][]{1985ApJS...58...39B,adkhikarietal14,moreetal15,diemer20}. While the splashback radius of galaxy clusters can also be inferred from the galaxy distribution of the clusters \citep[e.g.][]{moreetal16,changetal18,zurchermore19,shinetal19,murataetal20} and through weak lensing \citep{umetsudiemer17,contigianietal19}, this radial feature is much less pronounced than the edge radius identified in \citetalias{2020arXiv200311555T}. In addition, splashback detection via galaxy distributions is known to be sensitive to selection effects \citep{buschwhite17,sunayamamore19}. A similar bias may take place when measuring edge radii with velocity dispersions, though one benefits from the fact that cluster selection is done based on the galaxy density and not the velocity dispersion itself.  Consequently, we expect such biases to be reduced.  

Here, we treat the cluster edge radius in the velocity dispersion as a standard ruler which can be calibrated in simulations. Combined with the model of the phase space structure of clusters presented in \citetalias{2020arXiv200311555T}, we show that we are able to measure the angular diameter distance using the relative line-of-sight velocities of galaxies around clusters.  The structure of this paper is as follows. We describe our method in \cref{sec:method}. In \cref{sec:sdssResults} we demonstrate our method on current data and in \cref{sec:forecast}, and predict how our method will perform with near-future data. Our conclusions are presented in \cref{sec:conclusion}. Throughout the paper, unless otherwise stated, we assume a fiducial Flat $\Lambda$CDM cosmology with $h = 0.7$ and $\Omega_m = 0.3$. We emphasise that this paper is \emph{not} presenting a new measurement of the Hubble constant. Rather, our goal is simply to determine whether the method proposed here is sufficiently interesting to merit further investigation.

%%%%%%%%%%%%%%%%%%%%%%%%%%%%%%%%%%%%%%%%%%%%%%%%
%%%%%%%%%%%%%%%%%%%%%%%%%%%%%%%%%%%%%%%%%%%%%%%%
%%%%%%%%%%%%%%%%%%%%%%%%%%%%%%%%%%%%%%%%%%%%%%%%
%%%%%%%%%%%%%%%%%%%%%%%%%%%%%%%%%%%%%%%%%%%%%%%%
%%%%%%%%%%%%%%%%%%%%%%%%%%%%%%%%%%%%%%%%%%%%%%%%
%%%%%%%%%%%%%%%%%%%%%%%%%%%%%%%%%%%%%%%%%%%%%%%%
%%%%%%%%%%%%%%%%%%%%%%%%%%%%%%%%%%%%%%%%%%%%%%%%

\section{Methodology}
\label{sec:method}

\subsection{The Key Idea}

% Normally the next section describes the techniques the authors used.
% It is frequently split into subsections, such as Section~\ref{sec:maths} below.

\begin{figure}
    \centering
    \input{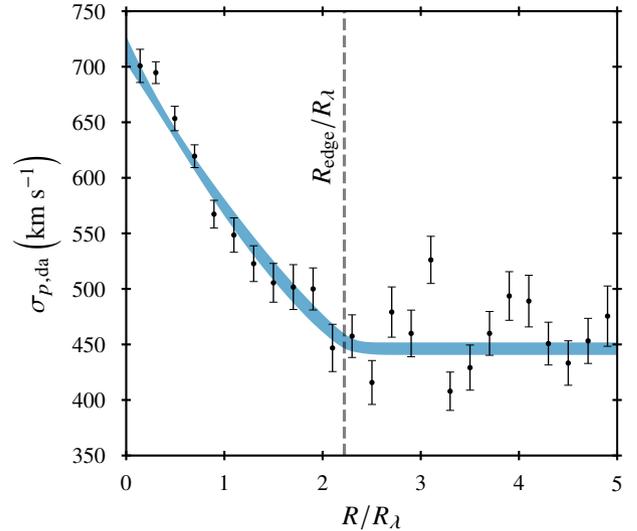}
    \caption{The radial dependence of the velocity dispersion of galaxies dynamically associated with SDSS \redmapper{} clusters. At small radii, the velocity dispersion decreases with increasing radius. At large radii, the velocity dispersion appears constant with radius. The boundary between these two regimes is the cluster ``edge radius'' which we seek to use as a standard ruler. Recreated from the right panel of Figure 2 of \citetalias{2020arXiv200311555T}.}
    \label{fig:sigmapx}
\end{figure}

\Cref{fig:sigmapx}, taken directly from \citetalias{2020arXiv200311555T}, shows the radial profile of the stacked line-of-sight velocity dispersion of galaxies relative to the central galaxy of SDSS \redmapper{} clusters. The points with error bars are the SDSS measurements, while the blue band corresponds to the best fit model of \citetalias{2020arXiv200311555T}. The key feature we would like to highlight is the obvious ``kink'' on this plot. This kink was interpreted in \citetalias{2020arXiv200311555T} as the spatial extent of galaxies orbiting \redmapper{} clusters, and we referred to this scale as the edge radius. This interpretation received theoretical support from \citetalias{2020arXiv200311557A}, which analysed the three dimensional analysis of substructure velocities around dark matter halos in a numerical simulation, and established the presence of a qualitatively similar feature in the three-dimensional velocity field. They also demonstrated that this edge radius is a simple re-scaling of the traditional splashback radius. In upcoming work, we will establish that the three dimensional ``edge radius'' observed in \citetalias{2020arXiv200311557A} does in fact exactly corresponds to the ``kink'' in the line-of-sight velocity dispersion profile shown in \cref{fig:sigmapx} (Aung et al, in preparation).

Our idea then is simple: it is relatively obvious that halos with larger line-of-sight velocity dispersions must also have larger edge radii. In other words, more massive halos occupy more space. We assume that we can use numerical simulations to calibrate a scaling relation between the velocity dispersion and edge radius of a halo, $\Redge = A\sigma_v^\alpha$.  Using spectroscopic data, we can empirically measure $\sigma_v^2$, which in turn determines the halo edge. At the same time, the same spectroscopic observations used to determine $\sigma_v$ allow us to measure the velocity dispersion profile as a function of angular separation $\theta$, thereby determining the angular scale $\theta_{\rm kink}$ corresponding to the ``kink'' feature in fig.~\ref{fig:sigmapx}. The distance to a galaxy clusters is then recovered via
\begin{equation}
    D_A(z_{\rm cluster}) = \frac{\Redge}{\theta_{\rm kink}} 
    = \frac{A\sigma_v^a}{\theta_{\rm kink}}
\end{equation}
where $D_A$ is the angular diameter distance, and $A$ and $\alpha$ are to be calibrated from simulations. With measurements of $D_A(z)$ at a variety of redshifts we can readily recover the Hubble parameter $h$.

In practice, things are a little more complicated: clusters do not have a single velocity dispersion: orbiting and infalling galaxies are kinematically distinct, and the line-of-sight velocity dispersion of both components varies with radius. Moreover, one must stack the signals of many clusters to achieve a high signal-to-noise measurement.   However, one can readily accommodate these additional difficulties using forward-modelling as detailed below.

%\it If one can calibrate the relation between halo size and the galaxy velocity dispersion, \rm we can use line-of-sight velocity dispersion measurements to infer the physical size of a halo. By comparing the physical size of the halo to the angular scale defined by the ``kink'' in \cref{fig:sigmapx}, we can infer the distance to the galaxy cluster in question. We detail below how such a measurement can be made in practice. 

\subsection{Full Forward Model}
%\subsection{Full Numerical Model}

Following \citetalias{2020arXiv200311555T} (albeit with slightly updated notation), we assume that the radial dependence of the line-of-sight velocity dispersion of \emph{orbiting} galaxies in a cluster is given by
\begin{equation}
    \sigma_{v,\orb}(R) = \frac{\sigma_{\orb}}{\sqrt{1+k R/\Redge}}.
    \label{eq:profile}
\end{equation}
In the above equation, $k$ is a shape parameter, which could be fit from the data, or which could be calibrated from numerical simulations. Here, we will conservatively assume that this shape parameter is fit from data rather than known a priori, but we will discuss the impact of such prior information throughout. The amplitude of the velocity dispersion profile for orbiting galaxies is governed by the parameter $\sigma_{\orb}$. We assume in turn that this parameter is related to the halo edge radius via a simple power-law,
\begin{equation}
    \label{eq:RedgePerfectPL}
    \Redge{} = R_\p \left(\frac{\sigma_{\orb}}{\overline{\sigma}_{\orb}}\right)^a \left(\frac{1 + z_{\rm cen}}{1 + z_\p}\right)^b ,
\end{equation}
where $\overline{\sigma}_{\orb}$ is a pivot scale set by the experimenter. Based on the results from \citetalias{2020arXiv200311555T}, we chose $\overline{\sigma}_{\orb}=\SI{788}{\kilo\meter\per\second}$. Note that we have included a possible redshift dependence in the relation between velocity dispersion and edge radius. We assume the parameters $R_\p$, $a$, and $b$ are calibrated from simulation. These parameters could in principle be cosmology-dependent. Our naive expectation is that any such dependence will be weak, as the dynamical structure of the halo should primarily reflect the halo's absolute mass. In practice, however, a dedicated calibration effort is clearly needed. For this work, our fiducial results will assume that $R_\p$, $a$, and $b$ are known, though we also consider how our results degrade as the uncertainty in these parameters increases. With these assumptions, the velocity dispersion profile of galaxy clusters is governed by either one or two parameters ($\sigma_{\orb}, k$), depending on whether we assume the shape of the profile is fixed a priori or not. 

When fitting survey data we are unable to fit each cluster individually. Instead, we consider a stack of galaxy clusters binned according to some mass proxy, e.g. the richness $\lambda$ of \redmapper{} clusters (though a similar analysis can be carried out for clusters stacked based on X-ray or SZ mass proxies). Here, we will assume the velocity dispersion of orbiting galaxies in a galaxy cluster is perfectly correlated with cluster richness, so that we can write
\begin{equation}
    \sigma_{\orb}(\lambda) = \sigma_{\p,\orb}\left( \frac{\lambda}{\lambda_\p} \right)^{\alpha_\orb} \left( \frac{1+z}{1+\zp} \right)^{\beta_\orb}
    \label{eq:scaling}
\end{equation}
Of course, in practice there will necessarily be scatter between the two mass tracers. Since our goal is to determine whether this type of analysis can in principle reach high precision, we will postpone the investigation of how scatter impacts our analyses to future work.

To summarise: \cref{eq:scaling} characterises the amplitude of the orbiting velocity dispersion profile as a function of richness. The edge radius of a cluster is a function of this orbiting velocity dispersion, and is given by \cref{eq:RedgePerfectPL}. The radial dependence of the velocity dispersion profile is given by \cref{eq:profile}, where the shape parameter $k$ can either be fit by the data or calibrated using numerical simulations. With this model in hand, the probability that an orbiting galaxy observed at a separation angle $\theta$ from the central galaxy of a cluster have line-of-sight velocity $v$ is given by a Gaussian of zero mean and velocity dispersion $\sigma_{v,\orb}(R)$ with $R=D_A\theta$. 
% DN: D_A as already defined above, so fine not to redefine this again here.
%where $D_A$ is the angular diameter distance to the cluster.

%We consider the simplest possible case, in which $D_A$ depends only on the Hubble constant parameter $h$, an approximation valid only at low redshifts. In this case, the model parameters that are allowed to vary in our analysis are $h$, $\sigma_{\p,\orb}$, $\alpha_\orb$, $\beta_\orb$, and the shape parameter $k$, for a total of \num{5} parameters. However, this model is still incomplete. A full model must account for the existence of in-falling galaxies, as well as projected galaxies along the line of sight. 

The parameters so far only take into account the orbiting galaxies, while in fact, infalling and background galaxies must also be accounted for in the model. The velocity probability distribution for each of these components are also modelled as Gaussians with radius-independent velocity dispersions. The velocity dispersions are assumed to scale with richness and redshift, leading to analogues of \cref{eq:scaling}:
\begin{eqnarray}
    \sigma_{\inf}(\lambda) & = \sigma_{\p,\inf}\left( \frac{\lambda}{\lambda_\p} \right)^{\alpha_\inf} \left( \frac{1+z}{1+\zp} \right)^{\beta_\inf} \\
     \sigma_{\los}(\lambda) & = \sigma_{\p,\los}\left( \frac{\lambda}{\lambda_\p} \right)^{\alpha_\los} \left( \frac{1+z}{1+\zp} \right)^{\beta_\los} 
\end{eqnarray}
The above consideration adds \num{6} parameters to our model, namely $\sigma_{\p,\inf}$ and $\sigma_{\p,\los}$, along with the corresponding richness and redshift slopes. 

Finally, we must also model the radial profile of the fraction of galaxies that are orbiting, in-falling, and line-of-sight projections, all as a function of radius. 
%Again, we rely on the model of \citetalias{2020arXiv200311555T}. 
The fraction of galaxies that are dynamically associated with the cluster (infalling or orbiting) is set to
\begin{equation}
    f_{\rm da}(R) = \begin{cases}
    1 + a_1(R/\Redge)\ \text{for $R\leq \Redge$} \\
    1+a_1+b_1(R/\Redge-1)\ \text{for $R\geq \Redge$} \end{cases} .
\end{equation}
This model differs from equation 14 of \citetalias{2020arXiv200311555T} in that we do not include the quadratic term $a_2 (R/\Redge)^2$ as $a_2$ was found to be consistent with zero. We also explicitly make the substitution $b_0 = 1 + a_1$ required for continuity at $R=\Redge$. The fraction of dynamically associated galaxies that are orbiting is set to
\begin{equation}
    f_\orb(R) = \begin{cases}
     c_0 + c_1(R/\Redge) + c_2(R/\Redge)^2\ \text{for $R\leq \Redge$} \\
    0\ \text{for $R\geq \Redge$} \end{cases} ,
\end{equation}
where we enforce that the equation is continuous at $\Redge$ by requiring that $c_0 = -(c_1 + c_2)$. These two radial profiles add an additional \num{4} free parameters ($a_1$, $b_1$, $c_1$, and $c_2$) that are to be fit from the data, bringing the total number of free parameters in the model to \num{15}. Our final model for the line-of-sight velocity for a galaxy at an angular separation $\theta$ from a galaxy cluster of richness $\lambda$ is
\begin{equation}
    P(v,\theta) = f_{\rm da}[f_{\orb}G_\orb(v) + (1-f_\orb) G_\inf(v)] + (1-f_{\rm da})G_\los(v),
\end{equation}
where $G_\orb$, $G_\inf$, and $G_\los$ are Gaussians with velocity dispersions $\sigma_\orb(D_A\theta|\lambda)$, $\sigma_\inf(\lambda)$ and $\sigma_\los(\lambda)$, respectively. By fitting this model to spectroscopic survey data, we are able to measure the Hubble parameter $h$.

\section{Prospects for Measuring \texorpdfstring{$h$}{h} with SDSS}
\label{sec:sdssResults}

To estimate the constraining power of this cluster and galaxy sample, we begin by assuming the parameters linking the orbiting velocity dispersion to the edge radius are perfectly calibrated. Based on the results of \citetalias{2020arXiv200311555T}, we set them to the values $R_\p=\SI{2.4}{\mega\parsec}$, $a=0.64$, and $b=-0.94$. Note that \citetalias{2020arXiv200311555T} assumed a fixed cosmology to recover these parameters, so our analysis is circular. That is, we are \emph{not} deriving any cosmological constraints. We are simply determining the precision with which we could measure $h$ if the parameters $R_\p$, $a$, and $b$ were known from simulations. 

We fit our model by sampling our \num{15}-dimensional parameter space with a Markov Chain Monte Carlo (MCMC) via \texttt{emcee}\footnote{\url{https://emcee.readthedocs.io/en/stable/}} \citep{2013PASP..125..306F}. We find $\sigma_h \approx \sigmahfid$, corresponding to a \SI{3}{\percent} measurement of the Hubble constant. Note that our forecast of $\sigma_h \approx \sigmahfid$ for an SDSS measurement is obtained while floating all \num{15} parameters in our model. In principle, numerical simulations could also fix or place priors on the parameters $a_1$, $b_1$, $c_1$, $c_2$, $\sigma_\inf$, $\alpha_\inf$, and $\beta_\inf$. When fixing these parameters in addition to $k$, we find $\sigma_h \approx \sigmahbest$, i.e. a \SI{1.2}{\percent} measurement of the Hubble constant with SDSS spectroscopy. We adopt the model in which all \num{15} parameters are allowed to float as our fiducial model, and use this as our basis for the DESI forecast.

\begin{figure}
    \centering
    \input{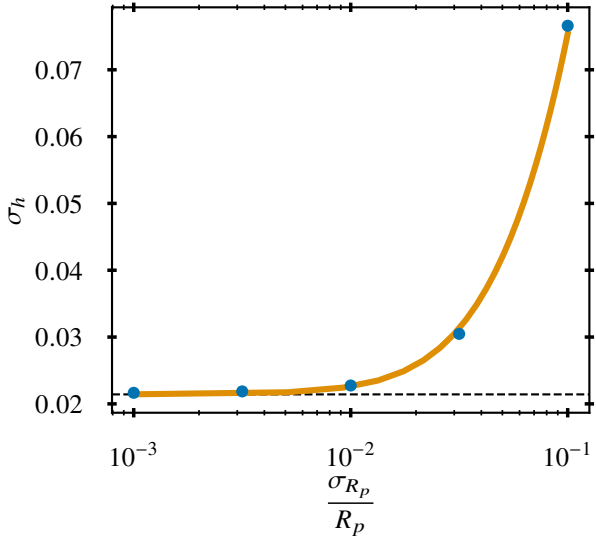}
    \caption{\label{fig:sigmahDeps} The uncertainty in $h$ as a function of the uncertainty in the calibration of the pivot radius (blue points). The orange line is the naive expectation from error propagation (see \cref{eq:err1}). The black dashed line is the fiducial estimate in the case of a perfect calibration.}
\end{figure}

In practice, our analysis will be sensitive to theoretical uncertainty in the input parameters $R_\p$, $a$, and $b$ linking the velocity dispersion to the cluster edge radius. To quantify the sensitivity of our measurement to this uncertainty, we repeat our measurement allowing for uncertainty in our calibration parameters. Specifically, we modify \cref{eq:RedgePerfectPL} by introducing three new parameters, $\Delta_\p$, $\Delta_a$, and $\Delta_b$, as per \cref{eq:RedgeUncertainty}.
\begin{equation}
    \label{eq:RedgeUncertainty}
    \Redge{} = R_\p \left(1 + \Delta_\p\right) \left(\frac{\sigma_{\mathrm{orb}}(\lambda, z)}{\overline{\sigma}_{\mathrm{orb}}}\right)^{a + \Delta_a} \left(\frac{1 + z_{\rm cen}}{1 + z_\p}\right)^{b + \Delta_b} .
\end{equation}
We consider varying each of these new parameters, one at a time, while holding the other two fixed. When varying the calibration parameters $\Delta$, we adopt a Gaussian prior for the parameter being varied. The prior has mean zero, with the standard deviation in $\Delta$ varied over the range \numrange{0.001}{0.1}. That is, the parameters $\Delta$ characterise the uncertainty with which the parameters $R_p$, $a$, and $b$ are known. We find that the posterior in $h$ is not sensitive to variations of these magnitudes in $a$ or $b$. For the pivot radius $R_p$, however, we do see that $\sigma_h$ increases with increasing calibration uncertainty as shown in \cref{fig:sigmahDeps}. The black dashed line in \cref{fig:sigmahDeps} is the fiducial uncertainty, and the blue points are the recovered uncertainty when allowing $\Delta_\p$ to vary. The orange line is our naive expectation due to error propagation,
\begin{equation}
    \sigma_h = \sqrt{\left(\frac{h_{\rm fid} \sigma_{R_\p}}{R_\p}\right)^2 + \sigma_{h, \mathrm{fid}}^2} .
    \label{eq:err1}
\end{equation}

Unsurprisingly, the quality of the recovered Hubble constant constraint is sensitive to uncertainties in the calibration of the relation between the halo edge and the halo velocity dispersion. In other words, enabling a percent level calibration of the Hubble constant using halo edges requires that halo edges be calibrated with better than percent level accuracy. 

%%%%%%%%%%%%%%%%%%%%%%%%%%%%%%%%%%%%%%%%%%%%%%%%
%%%%%%%%%%%%%%%%%%%%%%%%%%%%%%%%%%%%%%%%%%%%%%%%
%%%%%%%%%%%%%%%%%%%%%%%%%%%%%%%%%%%%%%%%%%%%%%%%
%%%%%%%%%%%%%%%%%%%%%%%%%%%%%%%%%%%%%%%%%%%%%%%%
%%%%%%%%%%%%%%%%%%%%%%%%%%%%%%%%%%%%%%%%%%%%%%%%
%%%%%%%%%%%%%%%%%%%%%%%%%%%%%%%%%%%%%%%%%%%%%%%%
%%%%%%%%%%%%%%%%%%%%%%%%%%%%%%%%%%%%%%%%%%%%%%%%

\section{DESI FORECAST}
\label{sec:forecast}

In our SDSS forecast, we only considered variations in the Hubble parameter $h$ in our analysis of SDSS data. By contrast, the DESI analysis will both reach higher precision and extend to higher redshifts, which will render the measurement sensitive to additional cosmological parameters. To determine the constraining power of our method in this case, we note that our measurement is intrinsically sensitive to $D_A$, which is itself proportional to $h^{-1}$. Consequently, we simply forecast the percent error for $h$ in DESI with all other cosmological parameters fixed for a narrow redshift bin, and then reinterpret this error as a percent uncertainty in $D_A$. By doing this across a grid of redshifts, we arrive at an array of predicted constraints on $D_A$ along this grid.  The resulting Hubble diagram can then be fit to recover the cosmological parameters of interest. 

\begin{figure}
    \centering
    \input{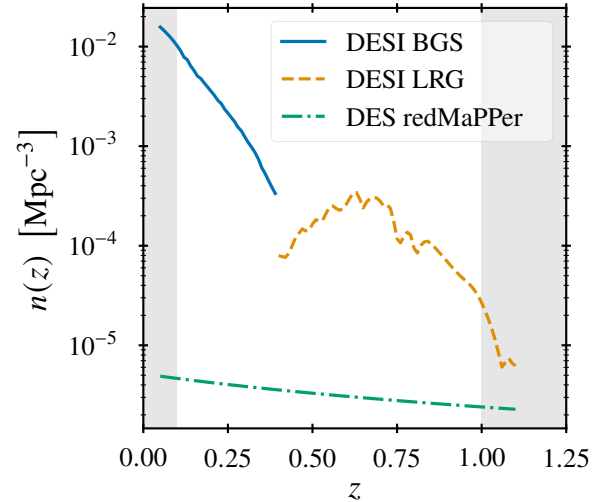}
    \caption{\label{fig:nz} The number densities as a function of redshift used for the DESI BGS (blue solid line), DESI LRG (orange dashed line), and DES \redmapper{} cluster (green dash-dotted line) samples. These number densities are used for the forecast, as described in \cref{sec:forecast}. Note that the DES number density is obtained by fitting a power law to the number of clusters as a function of redshift at lower redshifts, and then extending that power law to higher redshifts.}
\end{figure}

Our DESI forecast assumes that the \redmapper{} cluster catalogue is extended to redshift $z=1$ across the full footprint. This is a best case scenario, but as we show below, it ends up helping with the interpretation of our results. The cluster density out to redshift \num{1} is estimated using the Dark Energy Survey (DES) Year 1 \redmapper{} catalogue \citep{2019MNRAS.482.1352M}. Specifically, we fit a power law to the number of \redmapper{} clusters at low redshifts, and use this power law to estimate the number density at larger redshifts. We also estimate the galaxy density that is expected for the DESI survey from Figures 3.2 and 3.8 of \citet{2016arXiv161100036D}. We use the estimated density of the Bright Galaxy Sample (BGS) for redshifts $z \in \left[0.1, 0.4\right)$. At redshifts $z \geq 0.4$, we use the average of the estimates from COSMOS \citep{2009ApJ...690.1236I} and SDSS for the Luminous Red Galaxy (LRG) sample. The number density for both DESI galaxy samples as well as the cluster number density can be seen in \cref{fig:nz}.

In the absence of systematics, we expect that the measured error on $h$ will scale with the square-root of the number of galaxies used to estimate it, which we call $N_{\rm m}$, where the subscript ${\rm m}$ indicates that these spectroscopic galaxies must be matched to a cluster. The expected uncertainty in $h$ from a single redshift bin measurement, and assuming all other cosmological parameters are fixed, is given by
\begin{equation}
    \label{eq:sigmahForecast}
    \sigma_h^{\rm DESI}(z) = \sqrt{\frac{N_{\rm m}^{\rm SDSS}}{N_{\rm m}^{\rm DESI}(z)}} \sigma_h^{\rm SDSS} ,
\end{equation}
where $N_{\rm m}^x$ is the number of matched galaxies found for survey $x$ (SDSS or DESI). $\sigma_h^{\rm SDSS} = \sigmahfid$ is the measured error on $h$ in SDSS as estimated in the previous section. The number of galaxies matched to a cluster in a redshift bin should be proportional to the number of clusters in that redshift bin times the average number of galaxies per cluster:
\begin{equation}
    \label{eq:matchNum}
    N_m(z) = C N_{{\rm cl}}(z) N_{\sfrac{\rm g}{{\rm cl}}}(z) ,
\end{equation}
where $C$ is a proportionality constant.

The first term in \cref{eq:matchNum} can be found as the number density of clusters times the volume within the redshift bin:
\begin{equation}
    \label{eq:numCl}
    N_{{\rm cl}}(z) = n_{\cl}(z) V(z) .
\end{equation}
The second term in \cref{eq:matchNum} is the number density of galaxies times the volume within which a galaxy is considered a match to the cluster, $V_{\rm m}$:
\begin{equation}
    \label{eq:numGal}
    N_{\sfrac{\rm g}{\cl}}(z) = n_{\rm g}(z) V_{\rm m}(z) .
\end{equation}
For the match volume, we consider the same criteria as in 
\citetalias{2020arXiv200311555T} (section 3.1), assuming all clusters have a richness equal to the median richness of the \redmapper{} clusters. Our matching volume takes the form
\begin{equation}
    \label{eq:matchVol}
    V_{\rm m}(z) = \frac{25 \pi}{h^2} \left(\frac{\lambda(z)}{100}\right)^{0.4} \left[\chi(z + \Delta z) - \chi(z - \Delta z)\right] 
\end{equation}
where $2\Delta z$ is the width of the redshift bin of interest. The richness $\lambda(z)$ should in principle be the median richness within the redshift bin. In practice, however, the median richness of \redmapper{} clusters was found to be largely independent of redshift, so for simplicity we used the median richness of all clusters when estimating the matching volume.

\begin{figure*}
    \centering
    \includegraphics{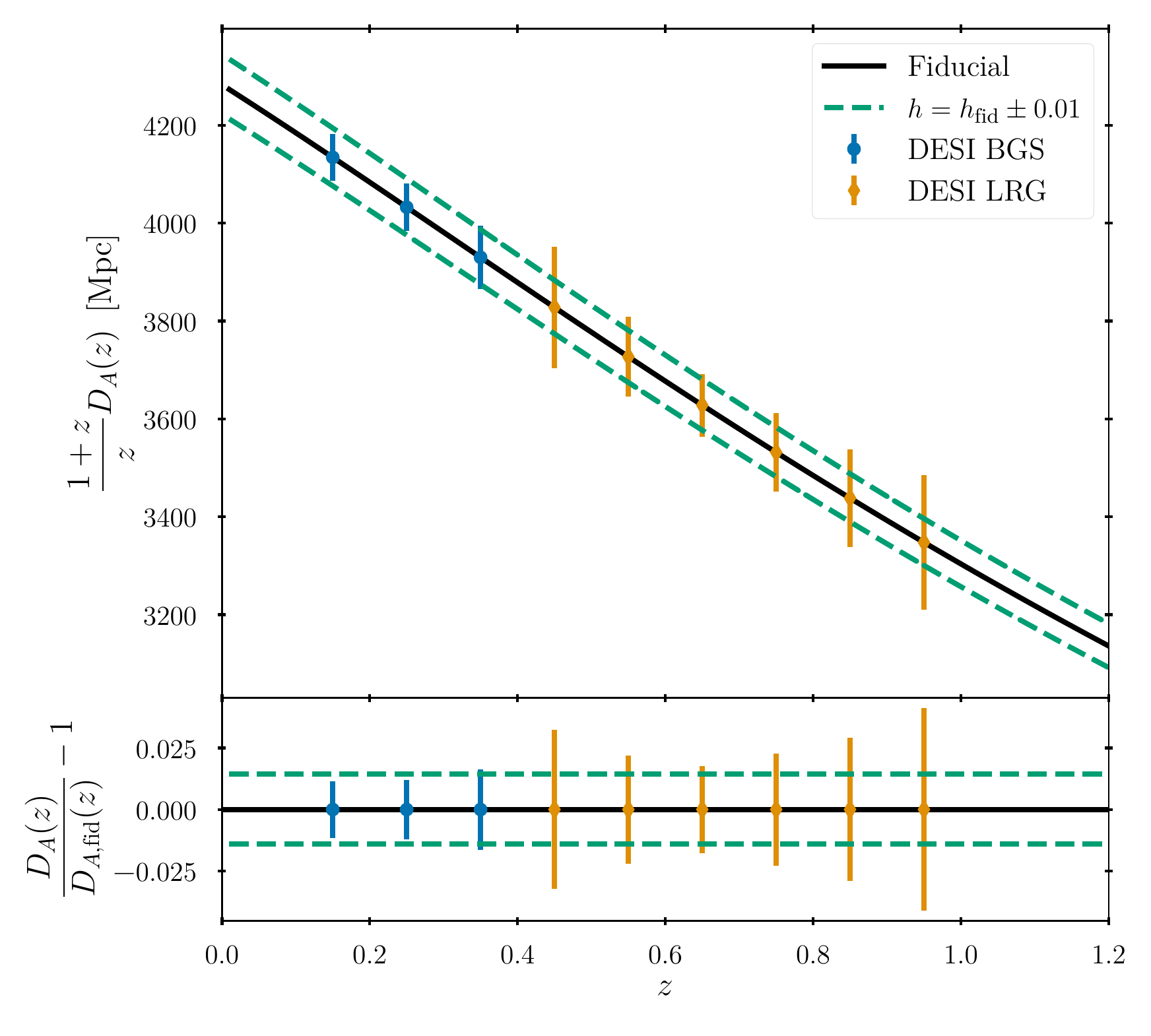}
    \caption{\label{fig:daForecast} Forecast of the error on the estimated angular diameter distance for DESI. The solid black line is the angular diameter distance for our fiducial cosmology. The dashed green lines are the angular diameter distance when the dimensionless Hubble constant is changed by \num{0.01} relative to the fiducial value. The blue error bars show the predicted uncertainty on the fiducial cosmology that can be measured by the DESI BGS sample assuming perfect calibration, and the orange error bars show the same but for the DESI LRG sample. The bottom panel shows the residual in the angular diameter distance relative to the fiducial cosmology.}
\end{figure*}

We determine the proportionality constant $C$ in \cref{eq:matchNum} using the SDSS data set. With these assumptions in hand, we can estimate the error $\sigma_h$ inferred from clusters in redshift bins of width $\pm \Delta z = 0.05$ between $z=0.1$ and $z=1$. As noted earlier, in practice, our measurement is sensitive to $D_A$, not $h$, so we re-interpret the predicted error in $h$ as a percent error on $D_A$ at a grid of redshifts $z=0.15, 0.25, \ldots, 0.95$. \Cref{fig:daForecast} shows $D_A(z)$ for our fiducial model, along with the predicted error on $D_A$. The green dashed lines show the change in $D_A$ when $h$ changes by \num{0.01}. The lower panel shows the residual from the fiducial angular diameter distance. 

\begin{figure}
    \centering
    \input{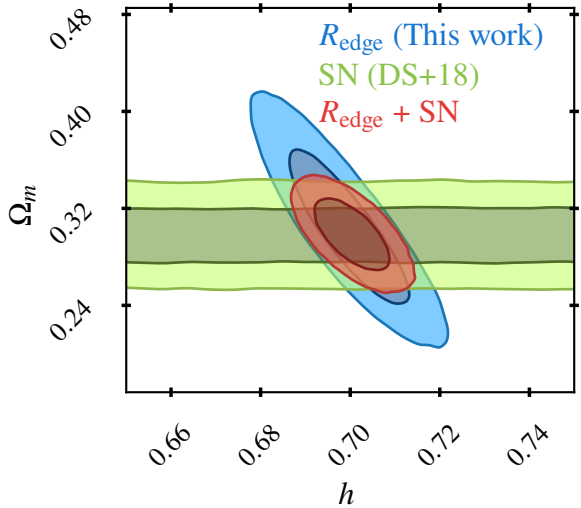}
    \caption{\label{fig:cosmoContours} Contours resulting from a 2-parameter ($h$ and $\Omega_m$) fit to the Hubble diagram data in \cref{fig:daForecast}. The blue contours are the result of fitting to the ``data'' given by the blue and orange points in \cref{fig:daForecast}. The green contours are the result of fitting the same parameters to the pantheon supernova sample \citep{2018ApJ...859..101S}. The red contours are the result of a combined analysis including both supernova data as well as the angular diameter distance ``data'' from \cref{fig:daForecast}.}
\end{figure}

We consider the cosmological constraints that could be derived from such a data set. To do so, we generate an artificial data vector comprised of the fiducial angular diameter distance and the errors from \cref{fig:daForecast}, and  fit for the cosmological parameters $h$ and $\Omega_{\rm m}$ assuming a flat $\Lambda$CDM model. The results are shown as the blue contours in \cref{fig:cosmoContours}. We find $\sigma_h \approx 0.009$, corresponding to a \SI{1.3}{\percent} measurement of $h$. The uncertainty in the recovered matter density parameters is large: $\sigma_{\Omega_m} \approx 0.042$. This can be compared to the constraint derived from the combined Pantheon supernova sample from \citet{2018ApJ...859..101S}, shown in \cref{fig:cosmoContours} as a green band ($\Omega_{\rm m}=0.298 \pm 0.022$). Adding the Pantheon data set as an external prior, our constraint on the Hubble parameter improves to $\sigma_h \approx 0.005$, or \SI{0.7}{\percent}. This is shown as the red contours in \cref{fig:cosmoContours}.

The large uncertainty in the recovered matter density parameter in our analysis demonstrates that the sensitivity of the proposed measurement to matter density and dark energy is quite limited: this measurement is simply not competitive with supernova measurements. This could in principle change if the cluster density is increased, and provided all shape parameters can be calibrated from simulations, but our analysis suggests such prospects are dim. Indeed, assuming all known nuisance parameters that can be calibrated from simulation are perfectly known, the error in $w_0$ for flat $w$CDM cosmology is $\sigma_w = 0.26$. By contrast, the predicted uncertainty in the Hubble parameter is exciting, and provides strong evidence that implementing the proposed measurement in the DESI data set, particularly within the context of the Bright Galaxy Survey, may result in highly competitive measurements of the Hubble constant. We emphasise that these constraints would be enabled by the relation between cluster size and galaxy velocity dispersion as calibrated from numerical simulations. As such, this standard-ruler measurement is calibrated exclusively through our understanding of gravity. While baryonic physics could in principle impact these predictions, the fact that the cluster radii are large suggests that baryonic effects on $\Redge$ will be small, though this will need to be quantitatively characterised in simulations. Given these features, our measurement is much more akin to a sound-horizon based measurement than the supernova measurements. That is to say, our calibration would ultimately be based on well understood physical processes, specifically, gravity.

%%%%%%%%%%%%%%%%%%%%%%%%%%%%%%%%%%%%%%%%%%%%%%%%
%%%%%%%%%%%%%%%%%%%%%%%%%%%%%%%%%%%%%%%%%%%%%%%%
%%%%%%%%%%%%%%%%%%%%%%%%%%%%%%%%%%%%%%%%%%%%%%%%
%%%%%%%%%%%%%%%%%%%%%%%%%%%%%%%%%%%%%%%%%%%%%%%%
%%%%%%%%%%%%%%%%%%%%%%%%%%%%%%%%%%%%%%%%%%%%%%%%
%%%%%%%%%%%%%%%%%%%%%%%%%%%%%%%%%%%%%%%%%%%%%%%%
%%%%%%%%%%%%%%%%%%%%%%%%%%%%%%%%%%%%%%%%%%%%%%%%

\section{Conclusions}
\label{sec:conclusion}

We have presented a novel method for measuring cosmological distances using the spatial extent of orbiting galaxies in halos as a standard ruler. Conceptually, a measurement of the galaxy velocity dispersion allows us to infer a halo's size. This size can in turn be detected as a specific angular scale, thereby enabling a measurement of the distance to the galaxy cluster. Critically, this is a first-principles physics-based calibration, and therefore does not rely on distance ladder measurements. At the same time, it shares no theoretical systematics with baryon acoustic oscillation and CMB measurements. 

To determine whether this measurement is interesting in principle, we adopted a fiducial model for the calibration between velocity dispersion and size, and proceeded to measure the Hubble constant, using  spectroscopic galaxies from SDSS DR14 and photometrically-selected galaxy clusters from the SDSS \redmapper{} algorithm. We emphasise that since we have \emph{assumed} the scaling relation between cluster velocity dispersion and halo radii, the central value of our Hubble constant measurement is meaningless. Our goal is only to forecast the precision of this newly proposed measurement. Our baseline uncertainty in the dimensionless Hubble constant as measured in the SDSS is $\sigma_h \approx \sigmahfid$ (\SI{3}{\percent}). Unsurprisingly, this precision is degraded if the calibration of the cluster edge radius in \cref{eq:RedgePerfectPL} is uncertain. Conversely, if all halo structure parameters are held fixed, the recovered uncertainty in the Hubble constant in the SDSS is $\sigma_h = \sigmahbest$ (\SI{1.2}{\percent}). 

We also performed a forecast of the constraining power of our method when using DESI data. We assumed a cluster catalogue extending to $z=1$, where the cluster density was estimated using the DES Y1 \redmapper{} cluster sample. Adopting a flat $\Lambda$CDM cosmology, we recover a \SI{1.3}{\percent} of the Hubble constant, which is dominated by low-redshift distance measurements. As we noted earlier, assuming that the \redmapper{} catalogue could be extended to redshift $z=1$ across the full DESI footprint is likely over-optimistic. Even with this assumption, however, we find that the corresponding error on the matter density is large: our proposed method is not competitive as a probe of dark energy. Thus, our proposed measurement works best as a low redshift technique for precisely measuring the Hubble constant. When we include a prior of $\Omega_{\rm m} = 0.298 \pm 0.022$ from the combined Pantheon supernova analysis of \citet{2018ApJ...859..101S}, our constraints improve to a \SI{0.7}{\percent} uncertainty on the Hubble constant. This level of precision more than suffices for unambiguously distinguishing between local and CMB-based values of the Hubble constant at high confidence.

There are, of course, additional obstacles that will need to be resolved in order to enable the proposed measurement, most notably characterising the impact of cluster selection effects on the recovered Hubble parameter. While such systematics may well increase the final delivered error substantially, there also exists the possibility of significant improvements relative to our forecast, particularly by calibrating additional halo structural parameters, e.g. the galaxy density profiles and the in-fall velocity dispersion profiles. Overall, our results provide strong motivation for aggressively pursuing the calibration and systematics tests necessary to enable the proposed Hubble constant measurement. If successful, such a program could provide smoking gun evidence either in favour or against the current Hubble tension problem. 

\section*{Acknowledgements}

ELW and ER are supported by the DOE grant DE-SC0015975. ER is also supported by DOE grant DE-SC0009913, and by NSF grant 2009401. HA and DN acknowledge support from Yale University. ER and DN also acknowledges funding from the Cottrell Scholar program of the Research Corporation for Science Advancement. 

Funding for SDSS-III has been provided by the Alfred P. Sloan Foundation, the Participating Institutions, the National Science Foundation, and the U.S. Department of Energy Office of Science. The SDSS-III web site is \url{http://www.sdss3.org/}.

SDSS-III is managed by the Astrophysical Research Consortium for the Participating Institutions of the SDSS-III Collaboration including the University of Arizona, the Brazilian Participation Group, Brookhaven National Laboratory, Carnegie Mellon University, University of Florida, the French Participation Group, the German Participation Group, Harvard University, the Instituto de Astrofisica de Canarias, the Michigan State/Notre Dame/JINA Participation Group, Johns Hopkins University, Lawrence Berkeley National Laboratory, Max Planck Institute for Astrophysics, Max Planck Institute for Extraterrestrial Physics, New Mexico State University, New York University, Ohio State University, Pennsylvania State University, University of Portsmouth, Princeton University, the Spanish Participation Group, University of Tokyo, University of Utah, Vanderbilt University, University of Virginia, University of Washington, and Yale University.

Funding for the Sloan Digital Sky Survey IV has been provided by the Alfred P. Sloan Foundation, the U.S. Department of Energy Office of Science, and the Participating Institutions. SDSS-IV acknowledges
support and resources from the Center for High-Performance Computing at
the University of Utah. The SDSS web site is \url{www.sdss.org}.

SDSS-IV is managed by the Astrophysical Research Consortium for the 
Participating Institutions of the SDSS Collaboration including the 
Brazilian Participation Group, the Carnegie Institution for Science, 
Carnegie Mellon University, the Chilean Participation Group, the French Participation Group, Harvard-Smithsonian Center for Astrophysics, 
Instituto de Astrof\'isica de Canarias, The Johns Hopkins University, Kavli Institute for the Physics and Mathematics of the Universe (IPMU) / 
University of Tokyo, the Korean Participation Group, Lawrence Berkeley National Laboratory, 
Leibniz Institut f\"ur Astrophysik Potsdam (AIP),  
Max-Planck-Institut f\"ur Astronomie (MPIA Heidelberg), 
Max-Planck-Institut f\"ur Astrophysik (MPA Garching), 
Max-Planck-Institut f\"ur Extraterrestrische Physik (MPE), 
National Astronomical Observatories of China, New Mexico State University, 
New York University, University of Notre Dame, 
Observat\'ario Nacional / MCTI, The Ohio State University, 
Pennsylvania State University, Shanghai Astronomical Observatory, 
United Kingdom Participation Group,
Universidad Nacional Aut\'onoma de M\'exico, University of Arizona, 
University of Colorado Boulder, University of Oxford, University of Portsmouth, 
University of Utah, University of Virginia, University of Washington, University of Wisconsin, 
Vanderbilt University, and Yale University.

Funding for the DES Projects has been provided by the U.S. Department of Energy, the U.S. National Science Foundation, the Ministry of Science and Education of Spain, 
the Science and Technology Facilities Council of the United Kingdom, the Higher Education Funding Council for England, the National Center for Supercomputing 
Applications at the University of Illinois at Urbana-Champaign, the Kavli Institute of Cosmological Physics at the University of Chicago, 
the Center for Cosmology and Astro-Particle Physics at the Ohio State University,
the Mitchell Institute for Fundamental Physics and Astronomy at Texas A\&M University, Financiadora de Estudos e Projetos, 
Funda{\c c}{\~a}o Carlos Chagas Filho de Amparo {\`a} Pesquisa do Estado do Rio de Janeiro, Conselho Nacional de Desenvolvimento Cient{\'i}fico e Tecnol{\'o}gico and 
the Minist{\'e}rio da Ci{\^e}ncia, Tecnologia e Inova{\c c}{\~a}o, the Deutsche Forschungsgemeinschaft and the Collaborating Institutions in the Dark Energy Survey. 

The Collaborating Institutions are Argonne National Laboratory, the University of California at Santa Cruz, the University of Cambridge, Centro de Investigaciones Energ{\'e}ticas, 
Medioambientales y Tecnol{\'o}gicas-Madrid, the University of Chicago, University College London, the DES-Brazil Consortium, the University of Edinburgh, 
the Eidgen{\"o}ssische Technische Hochschule (ETH) Z{\"u}rich, 
Fermi National Accelerator Laboratory, the University of Illinois at Urbana-Champaign, the Institut de Ci{\`e}ncies de l'Espai (IEEC/CSIC), 
the Institut de F{\'i}sica d'Altes Energies, Lawrence Berkeley National Laboratory, the Ludwig-Maximilians Universit{\"a}t M{\"u}nchen and the associated Excellence Cluster Universe, 
the University of Michigan, NFS's NOIRLab, the University of Nottingham, The Ohio State University, the University of Pennsylvania, the University of Portsmouth, 
SLAC National Accelerator Laboratory, Stanford University, the University of Sussex, Texas A\&M University, and the OzDES Membership Consortium.

Based in part on observations at Cerro Tololo Inter-American Observatory at NSF's NOIRLab (NOIRLab Prop. ID 2012B-0001; PI: J. Frieman), which is managed by the Association of Universities for Research in Astronomy (AURA) under a cooperative agreement with the National Science Foundation.

The DES data management system is supported by the National Science Foundation under Grant Numbers AST-1138766 and AST-1536171.
The DES participants from Spanish institutions are partially supported by MICINN under grants ESP2017-89838, PGC2018-094773, PGC2018-102021, SEV-2016-0588, SEV-2016-0597, and MDM-2015-0509, some of which include ERDF funds from the European Union. IFAE is partially funded by the CERCA program of the Generalitat de Catalunya.
Research leading to these results has received funding from the European Research
Council under the European Union's Seventh Framework Program (FP7/2007-2013) including ERC grant agreements 240672, 291329, and 306478.
We  acknowledge support from the Brazilian Instituto Nacional de Ci\^encia
e Tecnologia (INCT) e-Universe (CNPq grant 465376/2014-2).

This manuscript has been authored by Fermi Research Alliance, LLC under Contract No. DE-AC02-07CH11359 with the U.S. Department of Energy, Office of Science, Office of High Energy Physics.

This research made use of \texttt{Python}, \texttt{SciPy} \citep{2020NatMe..17..261V}, \texttt{NumPy} \citep{2011CSE....13b..22V}, IPython \citep{2007CSE.....9c..21P}, \texttt{Matplotlib} \citep{2007CSE.....9...90H}, and \texttt{ChainConsumer} \citep{2016JOSS....1...45H}. This research also made use of \texttt{Astropy},\footnote{\url{http://www.astropy.org}} a community-developed core Python package for Astronomy \citep{2013A&A...558A..33A,2018AJ....156..123A}.

%%%%%%%%%%%%%%%%%%%%%%%%%%%%%%%%%%%%%%%%%%%%%%%%%%
\section*{Data Availability}

The spectroscopic data underlying this article are available via query through the SDSS query service CasJobs at \url{http://skyserver.sdss.org/CasJobs/}. The SDSS \redmapper cluster catalogue is available at \url{http://risa.stanford.edu/redmapper}. The DES Y1 \redmapper{} catalogue is available for download at \url{https://des.ncsa.illinois.edu/releases/y1a1/key-catalogs/key-redmapper}.



\bibliographystyle{mnras}
\bibliography{references} % if your bibtex file is called example.bib

%\begin{thebibliography}{99}
%\end{thebibliography}

%%%%%%%%%%%%%%%%%%%%%%%%%%%%%%%%%%%%%%%%%%%%%%%%%%

%%%%%%%%%%%%%%%%% APPENDICES %%%%%%%%%%%%%%%%%%%%%

% \appendix

% \section{Some extra material}

% If you want to present additional material which would interrupt the flow of the main paper,
% it can be placed in an Appendix which appears after the list of references.

%%%%%%%%%%%%%%%%%%%%%%%%%%%%%%%%%%%%%%%%%%%%%%%%%%

% Don't change these lines
\bsp	% typesetting comment
\label{lastpage}

\end{document}